\documentclass[aps,pra,showpacs,twocolumn]{revtex4}
\usepackage{graphicx}
\begin{document}
\def\ket{\rangle}
\def\bra{\langle}
\def\w{\omega}
\def\W{\Omega}
\def\lr{\leftrightarrow}
\def\ud{\updownarrow}
\title{Modified entanglement purification scheme with doubly entangled photon state}

\author{Chuan Wang$^{a}$, Yu-Bo Sheng$^{b}$, Xi-Han Li$^{b}$, Fu-Guo Deng$^{c}$, Wei Zhang$^{d}$, Gui Lu Long$^{e}$}
\address{$^a$ School of Science and Key Laboratory of Optical Communication and Lightwave Technologies,
Beijing University of Posts and telecommunications, Beijing, 100876,
People's Republic of China\\
$^b$ Institute of Low Energy Nuclear Physics, and Department of
Material Science and Engineering,
Beijing Normal University, Beijing 100875, People's Republic of China\\
$^c$ Department of Physics, Applied Optics Beijing Area Major
Laboratory, Beijing Normal University, Beijing 100875, People's
Republic of China\\
$^d$ Department of Electronic Engineering, Tsinghua university, Beijing 100084, People's Republic of China\\
$^e$ Department of Physics, Tsinghua University, Beijing 100084,
People's Republic of China }

\date{\today }

\begin{abstract}
Recently Xiao et al. proposed a scheme for entanglement purification
based on doubly entangled photon states (Phys. Rev. A 77(2008)
042315). We modify  their scheme for improving the efficiency of
entanglement purification. This modified scheme contains two steps,
i.e., the bit-flip error correction and the entanglement
purification of phase-flip errors. All the photon pairs in the first
step can be kept as all the bit-flip errors are corrected. For
purifying the phase-flip errors, a wavelength conversion process is
needed. This scheme has the advantage of high efficiency and it
requires the original fidelity of the entangled state wanted fay
lower than other schemes, which makes it more feasible in a
practical application.
\end{abstract}
 \maketitle

\section{INTRODUCTION}

\label{ss1}

During the past decades, quantum entanglement presented many useful
properties in quantum information processing and transmission, such
as quantum key distribution \cite{ekert91,bbm92,LongLiu,CORE},
quantum dense coding \cite{bw}, quantum teleportation
\cite{teleportation}, and so on. In quantum communication, entangled
states are used to built quantum channel for information
transmission. However, in a practical condition the noise of the
channel will inevitably affect the entangled quantum states and even
make them be mixed states. The reduced entanglement of the quantum
systems will decrease the success probability of quantum
teleportation of an unknown state and even make a quantum key
distribution insecure. For accomplishing the task of secret quantum
communication, people should obtain some maximally entangled states
from a less-entanglement ensemble, called entanglement purification.
In 1996, Bennett et al. \cite{Bennett96} proposed the first protocol
for entanglement purification of Werner states with controlled-not
(CNOT) gates and bilateral rotations. In 2001, Pan et al.
\cite{Pan2} proposed an entanglement purification protocol with
polarization beam splitters (PBSs) and sophisticated single photon
detectors. In 2003, they experimentally demonstrated entanglement
purification of bit-flip error by using PBSs and four-path
coincidence photon counters \cite{Pan}. Also, entanglement
purification based on the parametric down-conversion (PDC)  source
was presented by Simon and Pan \cite{Simon}. Recently, Sheng, Deng
and Zhou \cite{Deng1} introduced a perfect protocol for entanglement
purification not only for PDC source but also for ideal source with
nonlinear optics. So far, entanglement purification have been widely
studied by many groups \cite{Murao,Horo,Deng2}.

Considering the novel idea of entanglement with multiple degrees of
freedom, many quantum communication protocols can be improved. For
instance, Aolita and Walborn \cite{aolita} proposed a quantum
communication protocol based on polarization and mode entangled
state, which has a high capacity. Barreiro et al. \cite{Barreiro}
had demonstrated a superdense coding experiment based on two degrees
of freedom of photons, which beats the channel capacity in quantum
communication. In 2005, Ravaro et al. \cite{ravaro} produced the
doubly entangled state with two degrees of freedom, i.e., the
frequency and the polarization of photons. They generated
two-photon states using semiconductor waveguides pumped by lasers.
The spontaneous parametric down conversion process generates a pair
of entangled photons with discrete frequencies.

Recently, Xiao et al. \cite{xiao} studied the properties of doubly
entangled photon state (DEPS) and proposed a entangled purification
protocol for DEPSs. This protocol can be realized with two steps,
i.e., the entanglement purification for bit-flip errors and that for
phase-flip errors. Two wavelength-division multiplexing (WDM)
devices are used in the first step to exclude the states with
bit-flip errors from the mixed ensemble. After the entanglement
purification for bit-flip error, the two parties select two states
from the remaining photon states and perform the PBS operations to
distinguish the states from those with phase-flip errors. However,
there is a problem in the second step as it requires that the two
photon pairs should be in  the same state, which means both of them
should be in the state $|\Phi^{+}\rangle$ or in the state
$|\Phi^{-}\rangle$. In this way, the second step cannot purify the
ensemble in a mixed state with the unit fidelity.

In this paper, we modify Xiao's entanglement purification protocol
for improving the success probability for entanglement purification
of bit-flip errors and decreasing phase-flip errors. In the first
step for the purification of bit-flip errors, we add two half wave
plates (HWPs) in the setups in Xiao's protocol. Just this
modification will make all the entangled states with or without
bit-flip errors kept as the bit-flip errors will be canceled by the
spatial modes and the HWPs. We complete the purification of
phase-flip errors in the second step following some ideas from the
protocol proposed by Pan et al. This protocol has the advantage of
high success probability and works more efficiently than Xiao's
protocol.

\section{Modified entanglement purification protocol based on DEPS}

\subsection{bit-flip error correction for a DEPS}

The DEPS generated in Ravaro's experiment can be described as
\cite{ravaro}:
\begin{equation}
\vert \Phi^{+}_{ab}\rangle
=\frac{1}{\sqrt{2}}(|H,\omega_{s}\rangle|H,\omega_{i}\rangle +
|V,\omega_{s'}\rangle|V,\omega_{i'}\rangle).
\end{equation}
Here H and V represent the horizontal and the vertical polarizations
of photons, respectively, and $\omega_{s(s')}$ and $\omega_{i(i')}$
correspond to the frequencies of entangled photons. Considering the
noisy channel transmission, we can not avoid the state to be
disturbed. The bit-flip errors or phase-flip errors will take place
on one of the particle or on both of the two particles. Then the
original state will be changed to
\begin{eqnarray}
\vert
\Phi^{\pm}_{ab}\rangle=\frac{1}{\sqrt{2}}(|H,\omega_{s}\rangle|H,\omega_{i}\rangle
\pm|V,\omega_{s'}\rangle|V,\omega_{i'}\rangle);\\
\vert
\Psi^{\pm}_{ab}\rangle=\frac{1}{\sqrt{2}}(|H,\omega_{s}\rangle|V,\omega_{i}\rangle
\pm|V,\omega_{s'}\rangle|H,\omega_{i'}\rangle);\\
\vert
\Gamma^{\pm}_{ab}\rangle=\frac{1}{\sqrt{2}}(|V,\omega_{s}\rangle|H,\omega_{i}\rangle
\pm|H,\omega_{s'}\rangle|V,\omega_{i'}\rangle);\\
\vert
\Upsilon^{\pm}_{ab}\rangle=\frac{1}{\sqrt{2}}(|V,\omega_{s}\rangle|V,\omega_{i}\rangle
\pm|H,\omega_{s'}\rangle|H,\omega_{i'}\rangle).
\end{eqnarray}
In this time, a pure state system transmitted may become a mixed
state ensemble described by the Werner state
\begin{eqnarray}
\rho & =&
F|\Phi^{+}_{ab}\rangle\langle\Phi^{+}_{ab}|+\frac{1-F}{7}|\Phi^{-}_{ab}\rangle\langle\Phi^{-}_{ab}|
+\frac{1-F}{7}|\Psi^{\pm}_{ab}\rangle\langle\Psi^{\pm}_{ab}| \nonumber \\
& + &
\frac{1-F}{7}|\Gamma^{\pm}_{ab}\rangle\langle\Gamma^{\pm}_{ab}|
+\frac{1-F}{7}|\Upsilon^{\pm}_{ab}\rangle\langle\Upsilon^{\pm}_{ab}|.
\end{eqnarray}
Here the coefficient
$F=\langle\Phi^{+}_{ab}|\rho|\Phi^{+}_{ab}\rangle$ is the fidelity
of initial state $|\Phi^{+}_{ab}\rangle$.

\begin{figure}
\begin{center}
\includegraphics[width=8cm,angle=0]{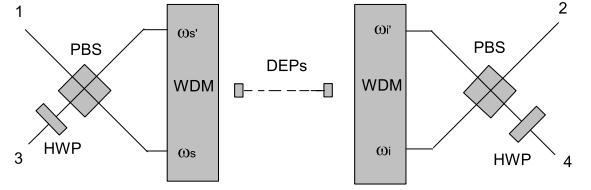}
\caption{ The setup for bit-flip error correction in the first
step. HWP and PBS represent half wave plate and polarizing beam
splitter, respectively. \label{f1}}
\end{center}
\end{figure}

Since a DEPS exhibits two degrees of freedom, we can utilize the
frequencies in our purification procedures to purify the
polarizations of photons. In our protocol, we exploit the
entanglement of frequencies to correct the bit-flip errors. Its
principle is shown in Fig.\ref{f1}.  When a DEPS enters the
device, one photon goes to the left wavelength-division
multiplexing device (WDM) and the other to the right one. Photons
with different frequencies can be distinguished by recording their
port information from WDMs. Since the two photons with different
frequencies leave their respective WDM in either
$\omega_{s}/\omega_{i}$ port or $\omega_{s'}/\omega_{i'}$ port,
each photon passes through a polarizing beam splitter and then
enters the port 1, 2, 3 or 4. Table \ref{t1} illustrates the
correspondence between the ports that each photon leaves the
device and the states.

\begin{table}
\caption{The correspondence between the ports and the
states}\label{t1}
\begin{center}
\begin{tabular}{ccccccc}
  \hline
  Triggered Port && Corresponding states \\
  \hline
  1,2 &&  $|\Phi^{\pm}_{ab}\rangle$ \\
  \hline
  1,4 &&  $|\Psi^{\pm}_{ab}\rangle$ \\
  \hline
  3,2 &&  $|\Gamma^{\pm}_{ab}\rangle$ \\
  \hline
  3,4 &&  $|\Upsilon^{\pm}_{ab}\rangle$ \\
  \hline
\end{tabular}
\end{center}
\end{table}

From Table  \ref{t1}, one can see that the state has no bit-flip
error if the two photons come out of the port 1 and the port 2,
respectively. If a photon comes out of the lower spatial mode (the
port 3 or the port 4), a bit-flip error takes place on it. We can
exploit the  two HWPs on the port 3 and the port 4  to correct the
bit-flip errors in the photon pair. For example, for the bit-flip
error state $|\Gamma^{+}_{ab}\rangle$, the two photons a and b
come out of the port 2 and the port 3, respectively. That is, a
bit-flip error takes place on the photon a coming out of the port
3. The HWP on the port 3 will accomplish the transformation $\vert
H\rangle $ $\leftrightarrow$ $\vert V\rangle$, which means that
the state of the photon pair becomes $|\Phi^{+}_{ab}\rangle$ after
the setup shown in Fig.1.

After the first step in our entanglement purification scheme, all
the states in the mixed ensemble are preserved and the bit-flip
errors are corrected. The DEPS system remains only the original
state $|\Phi^{+}_{ab}\rangle$ and the phase-flip error state
$|\Phi^{-}_{ab}\rangle$. That is, the initial Werner state becomes
to
\begin{eqnarray}
\rho'=\frac{4F+3}{7}|\Phi^{+}_{ab}\rangle\langle\Phi^{+}_{ab}|+\frac{4(1-F)}{7}|\Phi^{-}_{ab}\rangle\langle\Phi^{-}_{ab}|.
\end{eqnarray}

Compared with the Xiao's protocol, we only add two HWPs on the two
ports 3 and 4, but just this modification improves the efficiency
of the entanglement purification of bit-flip errors largely. In an
ideal condition, its efficiency for bit-flip error correction is
100\%.

\subsection{entanglement purification for phase-flip error}

In the second step, we purify the phase-flip error in the state
$\rho'$.  The key element before the second step is the wavelength
conversion process \cite{Takesue,Thew} which transforms the DEPSs
to traditional Bell states by a up-conversion process:
\begin{eqnarray}
|\Phi^{\pm}_{ab}\rangle\longrightarrow|\Phi^{\pm}_{Bell}\rangle=\frac{1}{\sqrt{2}}
(|H\rangle_{a}|H\rangle_{b}+|V\rangle_{a}|V\rangle_{b}).
\end{eqnarray}
This conversion process is performed in the first step after the
two WDMs. The wavelength conversion process consists of a WDM
coupler, a periodically poled lithium niobate (PPLN) waveguide and
a filter. One photon with the frequency $\omega_{s}$ pass through
the WDM as a signal light and pumped by a high power laser with a
certain wavelength. They are sent to PPLN waveguide and the
sum-frequency process generates the photon with the frequency
$\omega$ needed. The filter is used to filter the pump light. The
frequencies of both the a and b photons are performed by this
wavelength conversion to the same frequency, and then the DEPSs
are transformed to Bell states.

\begin{figure}
\begin{center}
\includegraphics[width=8cm,angle=0]{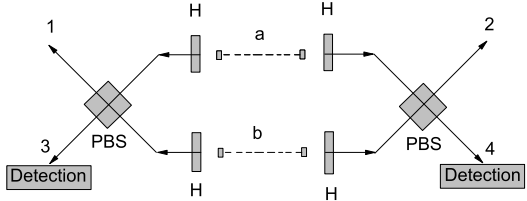}
\caption{ The setup for the second step in our purification
scheme, which is same as that in Ref.\cite{Pan2}. Here H and PBS
represent half wave plate and polarizing beam splitter,
respectively. \label{f2}}
\end{center}
\end{figure}

In the nonlocal entanglement purification procedures, the DEPS
with phase-flip errors can be further distilled with the ways in
other schemes existing, such as that with PBSs and sophisticated
single photon detectors \cite{Pan2}, shown in Fig.2. In detail,
the two parties, Alice and Bob, select two pairs of Bell states
(marked with $a_{1},a_{2}$ and $b_{1},b_{2}$) randomly and perform
a Hadamard operation on each of the photons. After these
operations, the phase-flip error state $|\Phi^{-}_{Bell}\rangle$
will be transformed into
$|\Psi^{+}_{Bell}\rangle=\frac{1}{\sqrt{2}}
(|H\rangle_{a}|V\rangle_{b}+|V\rangle_{a}|H\rangle_{b})$ while the
initial state $|\Phi^{+}_{Bell}\rangle$ remains unchanged. That
is, the state $\rho'$ becomes
\begin{equation}
\rho''=\frac{4F+3}{7}|\Phi^{+}_{ab}\rangle\langle\Phi^{+}_{ab}|+\frac{4(1-F)}{7}|\Psi^{+}_{ab}\rangle\langle\Psi^{+}_{ab}|.
\end{equation}
Alice and Bob choose the four-mode instances for their
entanglement purification, same as that in Ref. \cite{Pan2}. Alice
and Bob performs $\sigma_{x}=\{\vert
+x\rangle=\frac{1}{\sqrt{2}}(\vert H\rangle + \vert V\rangle),
\vert -x\rangle=\frac{1}{\sqrt{2}}(\vert H\rangle - \vert
V\rangle)\}$ measurement on the photons coming out of the port 3
and the port 4, shown in Fig.2. If their outcomes are
antiparallel, Alice and Bob need to take a phase-flip operation on
the photon pair coming from the up modes. After this purification
process, the state becomes
\begin{equation}
\rho'''=F'|\Phi^{+}_{ab}\rangle\langle\Phi^{+}_{ab}|+(1-F')|\Psi^{+}_{ab}\rangle\langle\Psi^{+}_{ab}|,
\end{equation}
where
\begin{equation}
F'=\frac{(4F+3)^2}{32F^2-8F+25}.
\end{equation}
When $\frac{4F+3}{7}$ is larger than $1/2$, $F'$ is larger than
$\frac{4F+3}{7}$ and the fidelity of the state $\vert
\Phi^+_{Bell}\rangle$ can be improved iteratively, which means our
entanglement purification scheme works at the case with the
original fidelity $F>\frac{1}{8}$, far lower than that in Ref.
\cite{Pan2}.

\section{Discussion and summary}

Compared to the traditional purification protocols, our scheme is
more efficient as the first step can correct all the bit-flip
errors. This good feature makes the original fidelity of the
entangled state wanted in our scheme much lower than others
\cite{Bennett96,Pan2,Pan,Simon,Deng1,Murao,Horo}. In the second
step for the purification of phase-flip errors, we can also use
the ways in other protocols to improve the fidelity, such as that
with nonlinear optics \cite{Deng1}. In experiment, the probability
of the wavelength conversion process is not unit, but it
approaches unit \cite{Takesue}. If the two parties possess some
cross-Kerr nonlinear media, it is possible for them to purify the
phase-flip errors without the wavelength conversion process.

In summary, we have improved the entangled purification scheme based
on  doubly entangled photon states. High fidelity DEPSs can be
achieved by two-step purification operations after the transmission
over a noisy channel. We have modified the purification setups using
WDMs, polarization beam splitters and high efficiency nonlinear
processes. In the first step operation, all the bit-flip error
states can be distinguished and corrected. The second step operation
can efficiently purify the phase-flip errors. Moreover, the original
fidelity of the entangled state wanted in our scheme is much lower
than others \cite{Bennett96,Pan2,Pan,Simon,Deng1,Murao,Horo}, which
makes this scheme more feasible in a practical application.

\section{ACKNOWLEDGMENTS}

This work is supported by the National Natural Science Foundation
of China under Grant No. 10604008.


\begin{thebibliography}{99}

\bibitem{ekert91} A.K. Ekert, Phys. Rev. Lett. 67 (1991) 661.

\bibitem{bbm92} C.H. Bennett, G. Brassard,  N.D. Mermin, Phys. Rev. Lett.
 68 (1992) 557 .

\bibitem{LongLiu} G.L. Long, X.S. Liu, Phys. Rev. A 65 (2002) 032302.

\bibitem{CORE} F.G. Deng, G.L. Long, Phys. Rev. A  68 (2003)  042315.


\bibitem{bw} C.H. Bennett, S.J. Wiesner, Phys. Rev. Lett.  69 (1992)
2881; X.S. Liu, G.L. Long, D.M. Tong,  L. Feng, Phys. Rev. A
 65 (2002) 022304; A. Grudka,  A. W\'ojcik, Phys. Rev. A
 66 (2002) 014301.


\bibitem{teleportation} C.H. Bennett, G. Brassard, C. Crepeau, R. Jozsa, A.
Peres,  W.K. Wootters, Phys. Rev. Lett.  70 (1993) 1895; F.G.
Deng, C.Y. Li, Y.S. Li, H.Y. Zhou,  Y. Wang, Phys. Rev. A
 72 (2005) 022338.


\bibitem{Bennett96} C.H. Bennett, G. Brassard, S. Popescu, B. Schumacher,
J.A. Smolin,  W.K. Wootters, Phys. Rev. Lett.  76 (1996) 722 .


\bibitem{Pan2} J.W. Pan, C. Simon,  A. Zellinger, Nature 410 (2001)
1067.

\bibitem{Pan} J.W. Pan, S. Gasparonl, R. Ursin, G. Weihs,  A.
Zellinger, Nature  423 (2003) 417.


\bibitem{Simon} C. Simon, J.W. Pan,  Phys. Rev. Lett.
 89 (2002) 257901.

\bibitem{Deng1} Y.B. Sheng, F.G. Deng,  H.Y. Zhou, Phys. Rev. A
 77 (2008) 042308.


\bibitem{Murao} M. Murao, M.B. Plenio, S. Popescu V. Vedral,  P.L. Knight, Phys. Rev. A  57 (1998) R4075 .

\bibitem{Horo} M. Horodecki, P. Horodecki, Phys. Rev. A 59 (1999) 4206.

\bibitem{Deng2} Y.B. Sheng, F.G. Deng,  H.Y. Zhou, Phys. Rev. A
 77 (2008) 062325.



\bibitem{aolita} L. Aolita, S.P. Walborn, Phys. Rev. Lett.  98 (2007)
100501.

\bibitem{Barreiro} J.T. Barreiro, T.C. Wei, P.G. Kwiat, Nature Physics
 4 (2008) 282.

\bibitem{ravaro} M. Ravaro, Y. Seurin, S. Ducci, G. Leo, V. Berger, A. De Rossi, G. Assanto, J. App. Phys.  98 (2005) 063103.

\bibitem{xiao} L. Xiao, C. Wang, W. Zhang, Y. D. Huang, J.D. Peng, G.L. Long, Phys. Rev. A  77 (2008)
042315.

\bibitem{Takesue} H. Takesue, E. Diamanti, T. Honjo, C. Langrock,
M.M. Fejer, K. Inoue, Y. Yamamoto, New. J. Phys.  7 (2005) 232.

\bibitem{Thew} R.T. Thew, S. Tanzilli, L. Krainer, S.C. Zeller, A. Rochas,
I. Rech, S. Cova, H. Zbinden, N. Gisin, New. J. Phys.
 8 (2006) 32.

\end{thebibliography}
\end{document}